\documentclass[aps,prl,twocolumn,groupedaddress,showpacs,amsmath,amssymb,floatfix,superscriptaddress]{revtex4-1}
\usepackage{multirow}
\usepackage{graphicx}
\usepackage{times}
\usepackage{color}
\usepackage{natbib}
\usepackage{array}
\newcolumntype{C}[1]{>{\centering\let\newline\\\arraybackslash\hspace{0pt}}m{#1}}

\begin{document}

\title{Search for bosonic super-weakly interacting massive particles at COSINE-100}
\author{G.~Adhikari}
\affiliation{Department of Physics and Wright Laboratory, Yale University, New Haven, CT 06520, USA}
\author{N.~Carlin}
\affiliation{Physics Institute, University of S\~{a}o Paulo, 05508-090, S\~{a}o Paulo, Brazil}
\author{J.~J.~Choi}
\affiliation{Department of Physics and Astronomy, Seoul National University, Seoul 08826, Republic of Korea} 
\affiliation{Center for Underground Physics, Institute for Basic Science (IBS), Daejeon 34126, Republic of Korea}
\author{S.~Choi}
\affiliation{Department of Physics and Astronomy, Seoul National University, Seoul 08826, Republic of Korea} 
\author{A.~C.~Ezeribe}
\affiliation{Department of Physics and Astronomy, University of Sheffield, Sheffield S3 7RH, United Kingdom}
\author{L.~E.~Fran{\c c}a}
\affiliation{Physics Institute, University of S\~{a}o Paulo, 05508-090, S\~{a}o Paulo, Brazil}
\author{C.~Ha}
\affiliation{Department of Physics, Chung-Ang University, Seoul 06973, Republic of Korea}
\author{I.~S.~Hahn}
\affiliation{Department of Science Education, Ewha Womans University, Seoul 03760, Republic of Korea} 
\affiliation{Center for Exotic Nuclear Studies, Institute for Basic Science (IBS), Daejeon 34126, Republic of Korea}
\affiliation{IBS School, University of Science and Technology (UST), Daejeon 34113, Republic of Korea}
\author{S.~J.~Hollick}
\affiliation{Department of Physics and Wright Laboratory, Yale University, New Haven, CT 06520, USA}
\author{E.~J.~Jeon}
\affiliation{Center for Underground Physics, Institute for Basic Science (IBS), Daejeon 34126, Republic of Korea}
\author{J.~H.~Jo}
\affiliation{Department of Physics and Wright Laboratory, Yale University, New Haven, CT 06520, USA}
\author{H.~W.~Joo}
\affiliation{Department of Physics and Astronomy, Seoul National University, Seoul 08826, Republic of Korea} 
\author{W.~G.~Kang}
\affiliation{Center for Underground Physics, Institute for Basic Science (IBS), Daejeon 34126, Republic of Korea}
\author{M.~Kauer}
\affiliation{Department of Physics and Wisconsin IceCube Particle Astrophysics Center, University of Wisconsin-Madison, Madison, WI 53706, USA}
\author{B.~H.~Kim}
\affiliation{Center for Underground Physics, Institute for Basic Science (IBS), Daejeon 34126, Republic of Korea}
\author{H.~J.~Kim}
\affiliation{Department of Physics, Kyungpook National University, Daegu 41566, Republic of Korea}
\author{J.~Kim}
\affiliation{Department of Physics, Chung-Ang University, Seoul 06973, Republic of Korea}
\author{K.~W.~Kim}
\affiliation{Center for Underground Physics, Institute for Basic Science (IBS), Daejeon 34126, Republic of Korea}
\author{S.~H.~Kim}
\affiliation{Center for Underground Physics, Institute for Basic Science (IBS), Daejeon 34126, Republic of Korea}
\author{S.~K.~Kim}
\affiliation{Department of Physics and Astronomy, Seoul National University, Seoul 08826, Republic of Korea}
\author{W.~K.~Kim}
\affiliation{IBS School, University of Science and Technology (UST), Daejeon 34113, Republic of Korea}
\affiliation{Center for Underground Physics, Institute for Basic Science (IBS), Daejeon 34126, Republic of Korea}
\author{Y.~D.~Kim}
\affiliation{Center for Underground Physics, Institute for Basic Science (IBS), Daejeon 34126, Republic of Korea}
\affiliation{Department of Physics, Sejong University, Seoul 05006, Republic of Korea}
\affiliation{IBS School, University of Science and Technology (UST), Daejeon 34113, Republic of Korea}
\author{Y.~H.~Kim}
\affiliation{Center for Underground Physics, Institute for Basic Science (IBS), Daejeon 34126, Republic of Korea}
\affiliation{Korea Research Institute of Standards and Science, Daejeon 34113, Republic of Korea}
\affiliation{IBS School, University of Science and Technology (UST), Daejeon 34113, Republic of Korea}
\author{Y.~J.~Ko}
\email{yjko@ibs.re.kr}
\affiliation{Center for Underground Physics, Institute for Basic Science (IBS), Daejeon 34126, Republic of Korea}
\author{D.~H.~Lee}
\affiliation{Department of Physics, Kyungpook National University, Daegu 41566, Republic of Korea}
\author{E.~K.~Lee}
\affiliation{Center for Underground Physics, Institute for Basic Science (IBS), Daejeon 34126, Republic of Korea}
\author{H.~Lee}
\affiliation{IBS School, University of Science and Technology (UST), Daejeon 34113, Republic of Korea}
\affiliation{Center for Underground Physics, Institute for Basic Science (IBS), Daejeon 34126, Republic of Korea}
\author{H.~S.~Lee}
\affiliation{Center for Underground Physics, Institute for Basic Science (IBS), Daejeon 34126, Republic of Korea}
\affiliation{IBS School, University of Science and Technology (UST), Daejeon 34113, Republic of Korea}
\author{H.~Y.~Lee}
\affiliation{Center for Underground Physics, Institute for Basic Science (IBS), Daejeon 34126, Republic of Korea}
\author{I.~S.~Lee}
\affiliation{Center for Underground Physics, Institute for Basic Science (IBS), Daejeon 34126, Republic of Korea}
\author{J.~Lee}
\affiliation{Center for Underground Physics, Institute for Basic Science (IBS), Daejeon 34126, Republic of Korea}
\author{J.~Y.~Lee}
\affiliation{Department of Physics, Kyungpook National University, Daegu 41566, Republic of Korea}
\author{M.~H.~Lee}
\affiliation{Center for Underground Physics, Institute for Basic Science (IBS), Daejeon 34126, Republic of Korea}
\affiliation{IBS School, University of Science and Technology (UST), Daejeon 34113, Republic of Korea}
\author{S.~H.~Lee}
\affiliation{IBS School, University of Science and Technology (UST), Daejeon 34113, Republic of Korea}
\affiliation{Center for Underground Physics, Institute for Basic Science (IBS), Daejeon 34126, Republic of Korea}
\author{S.~M.~Lee}
\affiliation{Department of Physics and Astronomy, Seoul National University, Seoul 08826, Republic of Korea} 
\author{Y.~J.~Lee}
\affiliation{Department of Physics, Chung-Ang University, Seoul 06973, Republic of Korea}
\author{D.~S.~Leonard}
\affiliation{Center for Underground Physics, Institute for Basic Science (IBS), Daejeon 34126, Republic of Korea}
\author{N.~T.~Luan}
\affiliation{Department of Physics, Kyungpook National University, Daegu 41566, Republic of Korea}
\author{B.~B.~Manzato}
\affiliation{Physics Institute, University of S\~{a}o Paulo, 05508-090, S\~{a}o Paulo, Brazil}
\author{R.~H.~Maruyama}
\affiliation{Department of Physics and Wright Laboratory, Yale University, New Haven, CT 06520, USA}
\author{R.~J.~Neal}
\affiliation{Department of Physics and Astronomy, University of Sheffield, Sheffield S3 7RH, United Kingdom}
\author{J.~A.~Nikkel}
\affiliation{Department of Physics and Wright Laboratory, Yale University, New Haven, CT 06520, USA}
\author{S.~L.~Olsen}
\affiliation{Center for Underground Physics, Institute for Basic Science (IBS), Daejeon 34126, Republic of Korea}
\author{B.~J.~Park}
\affiliation{IBS School, University of Science and Technology (UST), Daejeon 34113, Republic of Korea}
\affiliation{Center for Underground Physics, Institute for Basic Science (IBS), Daejeon 34126, Republic of Korea}
\author{H.~K.~Park}
\email{hyangkyu@korea.ac.kr}
\affiliation{Department of Accelerator Science, Korea University Sejong Campus, Sejong 30019, Republic of Korea}
\author{H.~S.~Park}
\affiliation{Korea Research Institute of Standards and Science, Daejeon 34113, Republic of Korea}
\author{K.~S.~Park}
\affiliation{Center for Underground Physics, Institute for Basic Science (IBS), Daejeon 34126, Republic of Korea}
\author{S.~D.~Park}
\affiliation{Department of Physics, Kyungpook National University, Daegu 41566, Republic of Korea}
\author{R.~L.~C.~Pitta}
\affiliation{Physics Institute, University of S\~{a}o Paulo, 05508-090, S\~{a}o Paulo, Brazil}
\author{H.~Prihtiadi}
\affiliation{Center for Underground Physics, Institute for Basic Science (IBS), Daejeon 34126, Republic of Korea}
\author{S.~J.~Ra}
\affiliation{Center for Underground Physics, Institute for Basic Science (IBS), Daejeon 34126, Republic of Korea}
\author{C.~Rott}
\affiliation{Department of Physics, Sungkyunkwan University, Suwon 16419, Republic of Korea}
\affiliation{Department of Physics and Astronomy, University of Utah, Salt Lake City, UT 84112, USA}
\author{K.~A.~Shin}
\affiliation{Center for Underground Physics, Institute for Basic Science (IBS), Daejeon 34126, Republic of Korea}
\author{D.~F.~F.~S. Cavalcante}
\affiliation{Physics Institute, University of S\~{a}o Paulo, 05508-090, S\~{a}o Paulo, Brazil}
\author{A.~Scarff}
\affiliation{Department of Physics and Astronomy, University of Sheffield, Sheffield S3 7RH, United Kingdom}
\author{N.~J.~C.~Spooner}
\affiliation{Department of Physics and Astronomy, University of Sheffield, Sheffield S3 7RH, United Kingdom}
\author{W.~G.~Thompson}
\affiliation{Department of Physics and Wright Laboratory, Yale University, New Haven, CT 06520, USA}
\author{L.~Yang}
\affiliation{Department of Physics, University of California San Diego, La Jolla, CA 92093, USA}
\author{G.~H.~Yu}
\affiliation{Department of Physics, Sungkyunkwan University, Suwon 16419, Republic of Korea}
\affiliation{Center for Underground Physics, Institute for Basic Science (IBS), Daejeon 34126, Republic of Korea}
\collaboration{COSINE-100 Collaboration}

\date{\today}

\begin{abstract}
We present results of a search for bosonic super-weakly interacting massive particles (BSW) as keV scale dark matter candidates that is based on an exposure of 97.7\,kg$\cdot$year from the COSINE experiment. In this search, we employ, for the first time, Compton-like as well as absorption processes for pseudoscalar and vector BSWs. No evidence for BSWs is found in the mass range from 10\,$\mathrm{keV/c}^2$ to 1\,$\mathrm{MeV/c}^2$, and we present the exclusion limits on the dimensionless coupling constants to electrons $g_{ae}$ for pseudoscalar and $\kappa$ for vector BSWs at 90\% confidence level. Our results show that these limits are improved by including the Compton-like process in masses of BSW, above $\mathcal{O}(100\,\mathrm{keV/c}^2)$.
\end{abstract}

\maketitle

Despite strong evidence for dark matter (DM) from a wide range of astrophysical and cosmological systems based on gravitational effects~\cite{Clowe:2006eq,planck_2016,bergstrom_2012,feng_2010}, its identity is one of the most puzzling questions in our understanding of the universe. One of the proposed candidates for DM is the weakly interacting massive particle (WIMP)~\cite{lee77,Goodman:1984dc,Vysotskii1977}, which could have been produced by a thermalized and freeze-out mechanism in the early universe. In the past decades, WIMPs with mass of $\mathcal{O}(100\,\mathrm{GeV/c}^2)$ have been extensively searched for by underground experiments but with no positive results~\cite{Battaglieri:2017aum}. This motivates various theoretical models that invoke low-mass dark-matter particles~\cite{Essig:2011nj,Hochberg:2016ajh,Knapen:2017xzo}.

In these models, the DM mass is considered to be greater than about 3\,keV/c$^2$ in order to avoid conflict with structure formation in the universe~\cite{markovic_2014}. Alternative DM candidates with mass scales ranging from keV to MeV, so-called bosonic super-weakly interacting massive particles~(BSW)~\cite{PhysRevD.78.115012,PhysRevLett.72.17,ELLIS1985175,PhysRevD.68.063504,Javier_Redondo_2009}, have been proposed. The BSW has experimental advantages compared to Fermionic super-WIMPs, such as the sterile neutrino and the gravitino, which are extremely difficult to detect. The BSWs could couple to the standard model particles as discussed in~\cite{PhysRevD.78.115012}, in which case they could be directly detected by the absorption process, in which an energy equal to its rest mass is deposited into a target atom in the detector.

\begin{figure*}[htb]
\begin{center}
\includegraphics[width=0.98\textwidth]{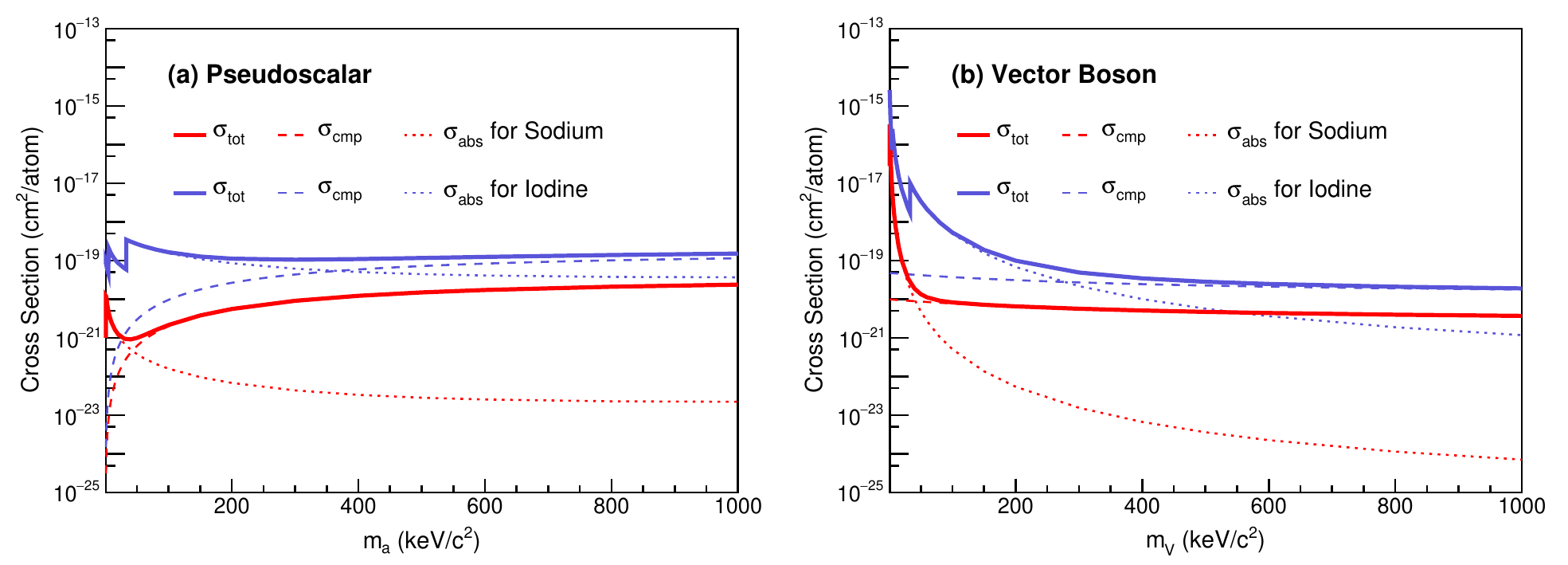} 
\caption{Cross sections for (a) pseudoscalar and (b) vector BSW with sodium (red) and iodine (blue) atoms. Dotted lines are the cross sections for the absorption process and dashed lines denote those for the Compton-like process. Solid thick lines show the total cross section for both processes.}
\label{fig_xc}
\end{center}
\end{figure*}

Results on BSW searches have been reported in the mass range of $\mathcal{O}(10-100\,\mathrm{keV/c}^2)$ by several experiments~\cite{PhysRevD.95.052006,PhysRevD.98.082004,PhysRevLett.118.161801,PhysRevD.101.052008,PhysRevD.102.072004,PhysRevLett.113.121301}, and recently the examined mass range has been extended to 1\,$\mathrm{MeV/c}^2$~\cite{PhysRevLett.125.011801}. These searches are based on the absorption process~\cite{PhysRevD.78.115012}. However, in the mass range above $\sim100\,\mathrm{keV/c}^2$, the cross section of the Compton-like process dominates over that of the absorption process as pointed out in~\cite{PhysRevD.104.083030}. Figure~\ref{fig_xc} shows the cross sections for BSW as a function of BSW mass for sodium and iodine atoms. The cross section for the Compton-like process for sodium (iodine) atoms and BSW masses above about $50\,\mathrm{keV/c}^2$ ($300\,\mathrm{keV/c}^2$) dominates that of the absorption process. Therefore, it is desirable to consider the Compton-like process, as well as the absorption process in a BSW search experiment. We have performed a search for the BSW in the mass range from $\mathrm{10\,keV/c}^2$ to $\mathrm{1\,MeV/c}^2$ that, for the first time, considers both the absorption and Compton-like processes.

The COSINE-100 detector~\cite{detector} is located in a water equivalent overburden of about 1800\,meters at the Yangyang underground laboratory in South Korea~\cite{kims1,kims2}. The active target of the detector consists of a 106-kg array of eight ultra-pure NaI(Tl) crystals. Photomultiplier tubes (PMTs) are attached to each end of each crystal to detect and amplify the scintillation signals from the crystal. The signals from the PMTs are recorded by a 500\,MHz flash analog-to-digital converter. The dynamic range was set to focus on energies of $\mathcal{O}$(keV) to detect scattering between atomic nuclei and WIMPs with masses on the $\mathcal{O}(100\,\mathrm{GeV/c}^2)$. However, by using additional channels that readout PMT dynode signals, the dynamic range for energies of $\mathcal{O}$(MeV) was recorded; thus, each crystal has two anode channels for low energy and two dynode channels for high energy. The dynamic range of the anode channel is from 1\,keV, the analysis threshold~\cite{eventselection}, to 70\,keV, whereas that of the dynode channel is from 70\,keV to 3000\,keV.

The crystal array is immersed in 2200\,liters of liquid scintillator (LS) that acts as an active shield~\cite{lsveto1,lsveto2}. The LS shields against the radiation coming from the outside of the crystal, as well as detects internal and external radiations. The LS container is a box with 1-cm-thick acrylic wall, surrounded by 3-cm-thick copper. The next layer is a 20-cm-thick lead shield against external radiation, and the outermost layer is a muon counter array of plastic scintillator panels. The muon counter array covers all directions of the detector and it is used to detect and veto cosmic-ray muon induced crystal signals~\cite{muon}. During the data taking, the detector environment such as radon level, temperature, {\it{etc}}., was continuously monitored~\cite{detector,monitoring}.

The data used for this analysis are from a 1.7\,year exposure recorded between October 2016 and July 2018. Three crystals were found to have high noise rates, so they were not used in this analysis, resulting in an effective exposure of 97.7\,kg$\cdot$year. Simulated data were modeled via the GEANT4 toolkit~\cite{geant4}. Since both anode and dynode channels are used for the data analysis, the energy range that was used to model the background was from 1\,keV to 3000\,keV. Scintillation events from NaI(Tl) crystals are classified into single-hit and multiple-hit events. Scintillation events are tagged as multiple-hit events if they occur in coincidence with LS or other crystals, and as single-hit events otherwise. Based on the multiplicity (single-hit and multiple-hit events) and the energy range (anode and dynode), the data are classified into four groups and modeled with simultaneous fits to each crystal~\cite{background2}. BSW masses larger than 1\,MeV are not considered in this analysis because they could decay into an $\rm e^+ + e^-$ pair with a lifetime that is too short to qualify for dark matter. Although the BSW mass-search-range only extends up to 1\,MeV/$\rm c^2$, the energy deposition to the crystals from background events is modeled up to 3\,MeV. Since the Compton-like process is dominant in the energy range of $\mathcal{O}$(100\,keV), both the Compton-like and absorption process are used in the simulation of the BSW signals.

\begin{figure*}[htb]
\begin{center}
\includegraphics[width=0.98\textwidth]{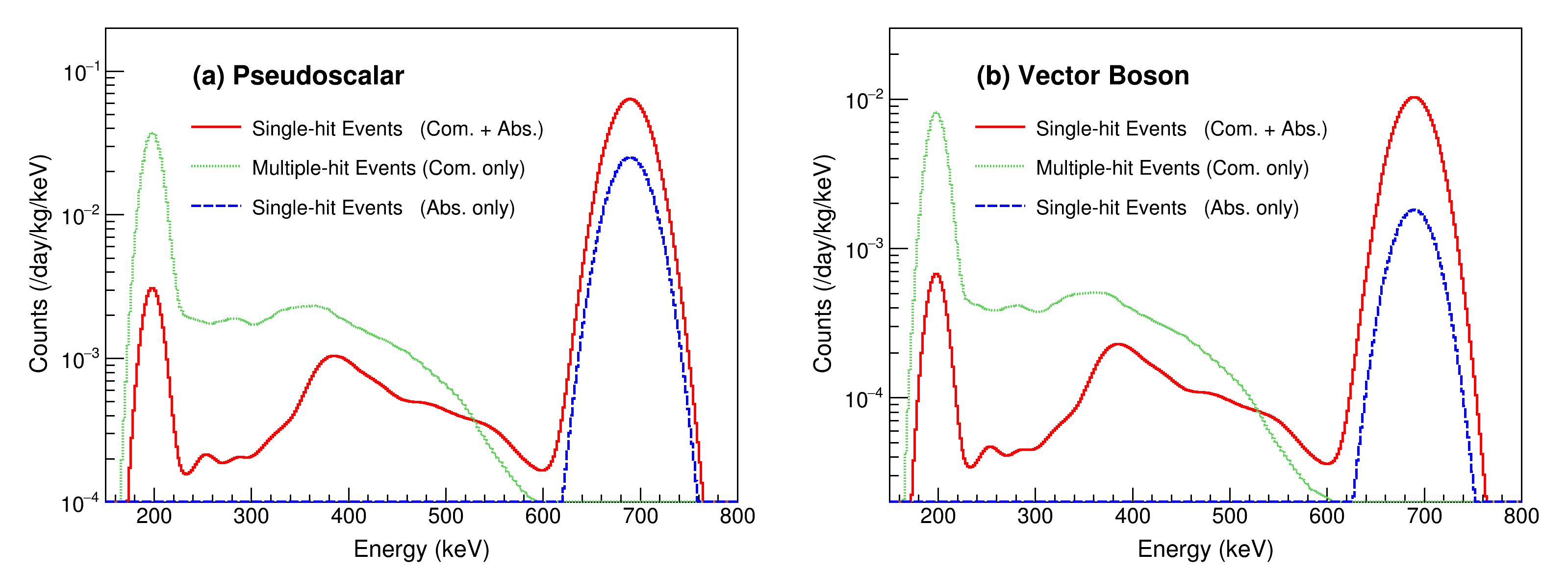} 
\caption{The expected energy spectra for a 690\,keV/c$^2$ BSW for a single crystal. Results for both a pseudoscalar BSW boson with $g_{ae} = 1$ and a vector BSW boson with $\kappa = 1$ are shown in (a) and (b), respectively. BSW events are generated for both the Compton-like and the absorption processes. The solid red lines represent the expected energy spectra for the single-hit events that are not accompanied by a detected signal in either the LS or any of other crystals. The dotted green lines show the energy spectra from the Compton-like energy deposition in either other crystals or the LS. The dashed blue lines are the expected energy spectra of BSW assuming only absorption process.}
\label{fig_sig}
\end{center}
\end{figure*}

The absorption of a BSW by an atom is similar to the photoelectric effect, with the photon energy $\omega \approx m_{a} (m_V)$, where $m_a$ ($m_V$) is the mass of pseudoscalar boson $a$ (vector boson $V$). Since the BSW is expected to be moving slowly, the energy transferred to the atom will approximately be equal to the BSW mass. The counting rate for the process can be expressed via the cross section for the photoelectric effect $\sigma_\mathrm{pe}(\omega)$. In the absorption process, an electron is emitted from the atom and the BSW mass is converted to electron kinetic energy. The counting rate for the pseudoscalar $a$ is related to a dimensionless coupling $g_{ae}$~\cite{PhysRevD.78.115012},
\begin{eqnarray}
R_a = \frac{1.2\times10^{19}}{A_\mathrm{Na} + A_\mathrm{I}}g_{ae}^2\left(\frac{m_a}{\mathrm{keV/c^2}}\right)\left(\frac{\sigma_\mathrm{pe}^\mathrm{sum}}{\mathrm{barn}}\right)\mathrm{d}^{-1}\mathrm{kg}^{-1},
\end{eqnarray}
where $A_\mathrm{Na}$ and $A_\mathrm{I}$ are atomic masses of sodium and iodine, respectively, and $\sigma_\mathrm{pe}^\mathrm{sum} = \sigma_\mathrm{pe}^\mathrm{Na} + \sigma_\mathrm{pe}^\mathrm{I}$ is the sum of cross sections for photoelectric effect on sodium and iodine atoms. In the case where the BSW is a vector boson $V$, the counting rate can be expressed as~\cite{PhysRevD.78.115012},
\begin{eqnarray}
R_V = \frac{4\times10^{23}}{A_\mathrm{Na} + A_\mathrm{I}}\frac{(e\kappa)^2}{4\pi\alpha}\left(\frac{\mathrm{keV/c^2}}{m_V}\right)\left(\frac{\sigma_\mathrm{pe}^\mathrm{sum}}{\mathrm{barn}}\right)\mathrm{d}^{-1}\mathrm{kg}^{-1},
\end{eqnarray}
where $e$ is the electron charge, $\kappa$ is the kinetic mixing parameter for the vector boson $V$ with the electromagnetic field, and $\alpha$ is the fine structure constant. In the absorption process, only the emitted electron contributes its energy deposition equivalent to the BSW mass into the crystal providing single-hit events.

In order to obtain the counting rate for the Compton-like process of BSW with electrons in the NaI crystals, we use a calculation given in~\cite{PhysRevD.104.083030}. In this process, both an electron and a photon are emitted via the interaction of BSW with an electron in the atom. The electron recoil energy ($T_e$) and the emitted photon energy ($E_\gamma$) are well determined by $T_e = m_{a,V}^2/[2(m_e +m_{a,V})]$ and $E_\gamma=m_{a,V} - T_e$ for a slowly moving BSW~\cite{Hochberg:2021zrf} where $m_e$ is the electron mass. The electron recoil energy is fully absorbed by the crystal. However, the photon can deposit a part of its energy into the crystal, escape out of the crystal, and leave its energy in either the LS or other crystals, which would produce a multiple-hit event.

BSW signal events for the absorption and Compton-like processes in the mass range from 10\,$\rm keV/c^2$ to 1000\,$\rm keV/c^2$ are simulated for bins smaller than the energy resolution and passed the through the COSINE-100 detector simulation, taking into account different detector responses for electrons and photons. Figure~\ref{fig_sig} shows the simulated energy distribution of the BSW signal for both processes for a BSW mass of 690\,keV/c$^2$ in a single crystal. The single-hit events from the absorption process show a peak at 690\,keV corresponding to the BSW mass. On the other hand, the Compton-like process contributes for both single-hit and multiple-hit events as shown in the figure. In the single-hit events from the Compton-like process, the full energy deposition from both the photon and the electron in a single crystal produces a 690\,keV peak, while the energy deposition from only the electron with no detectable energy deposition in either the LS or the other crystals produces a 200\,keV peak. In the multiple-hit events from the Compton-like process, electron produces a 200\,keV signal to a single crystal, while photon can deposit some energy in the same crystal or some in the other crystals or the LS.

\begin{figure*}[htb]
\begin{center}
\includegraphics[width=0.98\textwidth]{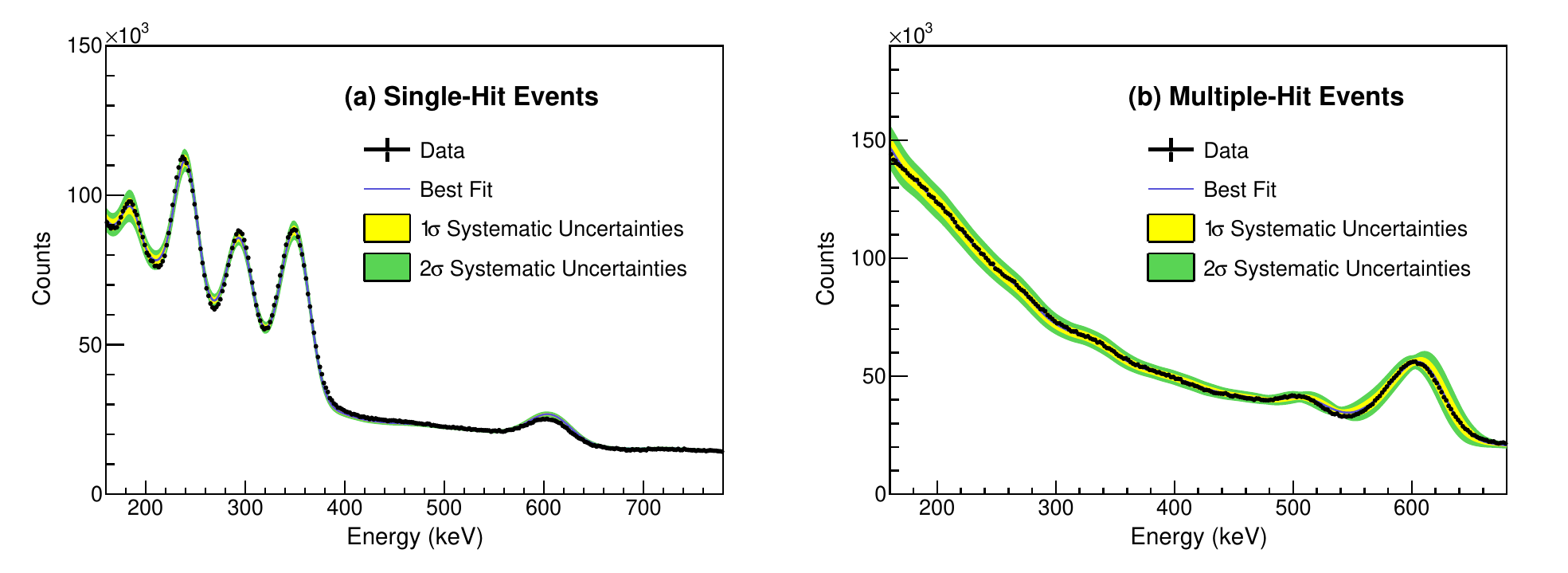} 
\caption{An example of extraction analysis results for a 690\,keV/c$^2$ pseudoscalar BSW. Black dots are summed data spectra of the five crystals. The best fit (blue) is presented with 1$\sigma$ (yellow) and 2$\sigma$ (green) systematic uncertainties. The (a) single-hit and (b) multiple-hit channels are fitted simultaneously.}
\label{fig_fit}
\end{center}
\end{figure*}

The BSW signals were simultaneously extracted from the measured energy spectra of the five crystals, and a Bayesian method was used. The posterior probability density function (PDF) for the BSW signal is described as
\begin{eqnarray}
P(\theta|\vec{M}) = \frac{\mathcal{L}(\vec{M}|\theta)\pi(\theta)}{\int\mathcal{L}(\vec{M}|\theta)\pi(\theta)d\theta},
\end{eqnarray}
where $\theta$ denotes $g_{ae}$ and $\kappa$ for pseudoscalar and vector BSW, respectively, which determines the signal strength. For the prior probability $\pi(\theta)$, a Heaviside step function was selected. The likelihood function $\mathcal{L}(\vec{M}|\theta)$ is marginalized to take into account the impact of variation of the energy resolution and scale, the event selection, and the background activities including the location of external radioactive sources, which are controlled by Gaussian constraints with their systematic uncertainties,
\begin{eqnarray}
\mathcal{L}(\vec{M}|\theta) = \int \mathcal{L}(\vec{M}|\theta,\vec{\alpha}) \pi(\vec{\alpha})\,d\vec{\alpha},
\end{eqnarray}
where $\vec{\alpha}$ denotes the nuisance parameters corresponding to systematic uncertainties, and $\pi(\vec{\alpha})$ denotes the Gaussian constraints. In order to marginalize the likelihood function, the Markov Chain Monte Carlo~\cite{mcmc1,mcmc2} is implemented through the Metropolis-Hastings algorithm~\cite{metropolis1,metropolis2}.

Figure~\ref{fig_fit} shows, as an example, the fit results for an assumed pseudoscalar BSW mass of 690 keV/c$^2$. Extraction of the pseudoscalar BSW signals generated by both processes was performed simultaneously on the single-hit and multiple-hit spectra for the five crystals. The spectra with best-fit values obtained from the posteriors of the nuisance parameters controlled by the Gaussian constraints are shown by blue lines. Similarly, the 1$\sigma$ and the 2$\sigma$ uncertainty of each parameter obtained from the posterior was propagated to form the systematic uncertainty bands. One can see that the data agree well with the fitted model within the systematic uncertainty band. A raster scan was performed in this way for BSW masses ranging from 10\,keV/c$^2$ to 1000\,keV/c$^2$.

\begin{figure*}[htb]
\begin{center}
\includegraphics[width=0.98\textwidth]{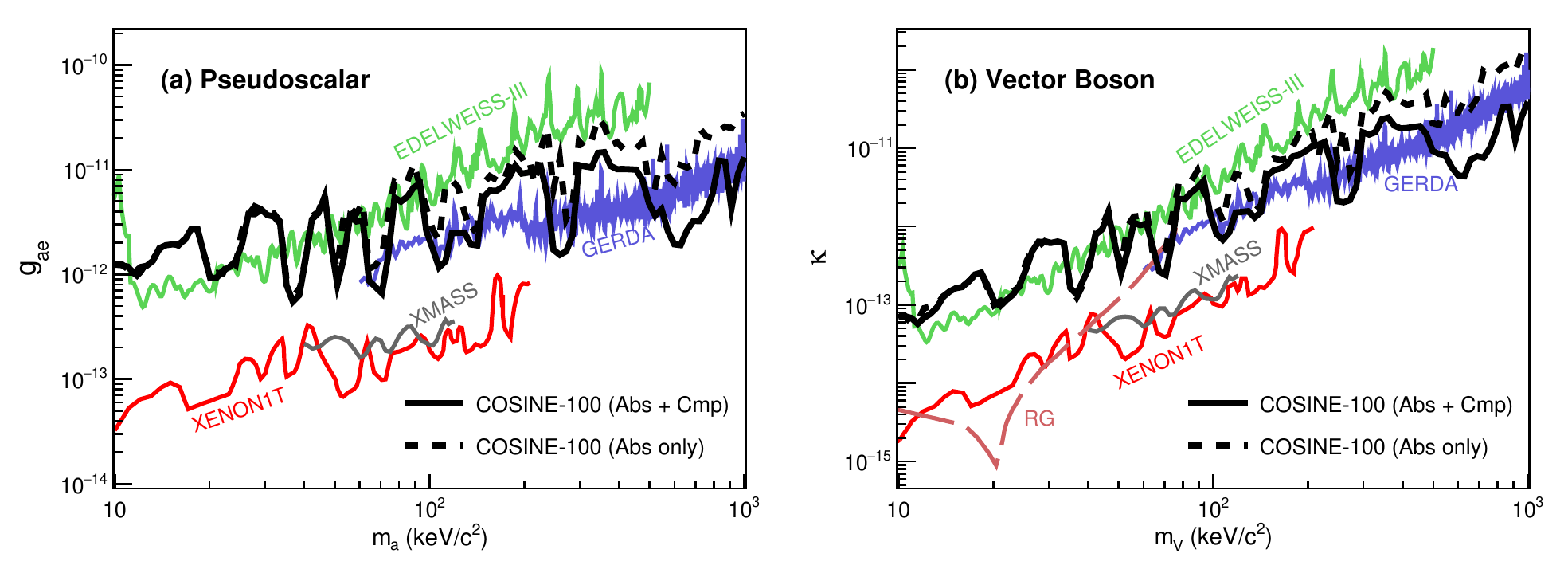} 
\caption{Exclusion limits (90\% C.L.) on the couplings for (a) pseudoscalar and (b) vector boson, as a function of mass. The solid black lines include Compton-like process as well as absorption process while the dashed black lines only consider the absorption process. Limits from other experiments or astrophysical constraint (red giant; RG) are also shown~\cite{PhysRevD.98.082004,PhysRevLett.125.011801,PhysRevD.102.072004,PhysRevLett.113.121301,rgstar}.}
\label{fig_limit}
\end{center}
\end{figure*}

There is no strong evidence for a non-zero signal posterior PDF for any BSW mass in the [10,~1000]\,$\mathrm{keV/c}^2$ range. Thus, exclusion limits on $g_{ae}$ and $\kappa$ at 90\% C.L. are determined from the posteriors. Figure~\ref{fig_limit} shows the exclusion limit curves for pseudoscalar and vector BSW. The black solid lines are the exclusion limits of COSINE-100 data for both processes while the black dashed lines show the limits for only the absorption process. The extraction limits including the Compton-like process provide better sensitivity than those that are based on the absorption process alone; this is especially the case for BSW masses above $\mathcal{O}$(100\,keV/c$^2$); the dimensionless couplings for pseudoscalar (vector) BSW to electron, $g_{ae}$ ($\kappa$), is improved up to 7.4 (12.9).

In summary, we performed a search for pseudoscalar and vector bosons of the BSW using 97.7-kg$\cdot$year COSINE-100 data in the BSW mass range from 10\,keV/c$^2$ to 1000\,keV/c$^2$. In this search, we included, for the first time, the Compton-like process, as well as the absorption process. There is no significant signal observed in this search, and we set constraints on the dimensionless couplings of pseudoscalar BSW and vector BSW to electrons, $g_{ae}$ and $\kappa$, respectively. By including the Compton-like process, the exclusion limits are improved in BSW masses above $\mathcal{O}$(100\,keV/c$^2$).

We thank the Korea Hydro and Nuclear Power (KHNP) Company for providing underground laboratory space at Yangyang and the IBS Research Solution Center (RSC) for providing high performance computing resources. 
This work is supported by:  the Institute for Basic Science (IBS) under project code IBS-R016-A1, NRF-2021R1A2C3010989,  NRF-2021R1I1A3041453, 
NRF-2021R1A2C1013761, NFEC-2019R1A6C1010027 and Korea University Grant, Republic of Korea;
NSF Grants No. PHY-1913742, DGE-1122492, WIPAC, the Wisconsin Alumni Research Foundation, United States; 
STFC Grant ST/N000277/1 and ST/K001337/1, United Kingdom;
Grants No. 2021/06843-1, No. 2022/12002-7 and No. 2022/13293-5 FAPESP, CAPES Finance Code 001, CNPq 303122/2020-0, Brazil.

\bibliography{dm}

\end{document}